\documentclass[10pt, a4paper]{article}

\usepackage{amsmath,amsfonts,amssymb}
\usepackage{graphicx}
\usepackage{upgreek} 

\usepackage[backend=biber,sorting=none,style=ieee,
doi=false,isbn=false,url=false,eprint=false]{biblatex}
\usepackage{authblk} 

\newcommand{\authortextsize}{\normalsize}
\newcommand{\affiltextsize}{\footnotesize}

\usepackage[font=small]{caption} 

\usepackage{titlesec} 
\titleformat*{\section}{\large\bfseries} 
\titleformat*{\subsection}{\normalsize\bfseries}

\title{\LARGE Adjustable spatial filter for optimal free-space quantum communication round the clock}

\author[1,2]{\authortextsize Andrej Kržič\footnote{Andrej.Krzic@iof.fraunhofer.de}}
\author[1]{\authortextsize Nico Döll}
\author[1]{\authortextsize Uday Chandrashekara}
\author[1,2]{\authortextsize Christopher Spiess}
\author[1,3]{\authortextsize Fabian Steinlechner\footnote{Fabian.Steinlechner@iof.fraunhofer.de}}
\affil[1]{\affiltextsize Fraunhofer Institute for Applied Optics and Precision Engineering, Albert-Einstein-Str. 7, 07745 Jena, Germany}
\affil[2]{\affiltextsize Friedrich Schiller University Jena, Faculty of Physics and Astronomy, Max-Wien-Platz 1, 07743 Jena, Germany}
\affil[3]{\affiltextsize Abbe Center of Photonics, Friedrich Schiller University Jena, Albert-Einstein-Str. 6, 07745 Jena, Germany}

\date{}

\begin{document}

\maketitle




\begin{abstract}
Free-space quantum communication in daylight relies crucially on spatial filtering. The optimal filter configuration, however, depends on ever-changing link conditions. To this end, we developed an adjustable spatial filter that can be used to change the system field of view on the fly. We demonstrate its use in quantum key distribution over a 1.7-km free-space link. Furthermore, we compare it to filtering with multi-mode fibre coupling. Finally, we extrapolate our results to a broader range of realistic link conditions and show that active field-of-view optimization has the potential to substantially improve the overall secure key output of the system.
\end{abstract}

\section{Introduction}
Quantum communication with light is a rapidly developing field \cite{Krenn.2016}. Its most technologically mature form is quantum key distribution (QKD) \cite{Pirandola.2020}. On the path towards large-scale QKD networks \cite{Zhang.2018}, the research focus has been increasingly moving from fundamental towards the more practical aspects.

One of the remaining practical challenges in quantum communication over free-space channels is operation in daylight \cite{Gruneisen.2015}. During daytime, a vast amount of sunlight enters the receiver telescopes and may end up completely masking the weak quantum signal upon detection. In order to reduce this background noise to acceptable levels, strong filtering is necessary in all the available degrees of freedom. Implementation of temporal filtering is relatively straightforward with well-established methods. Similarly, spectral filtering can be effortlessly implemented with off-the-shelf components. Efficient spatial filtering, on the other hand, remains a very challenging task due to atmospheric turbulence, which randomly distorts the signal wavefront at the receiver on millisecond time scales \cite{Andrews.2005}. This makes the light very difficult to focus through a pinhole or couple into an optical fibre. The majority of free-space quantum communication experiments to date were thus limited to low ambient light conditions, such as in nighttime.

Nevertheless, several proof-of-principle QKD experiments in daylight have been reported over terrestrial free-space links. The first successful demonstration was performed in 2000 with a faint pulse source, using multi-mode (MM) fibre coupling as a means of spatial filtering \cite{Hughes.2000}. The first entanglement-based implementation followed in 2009, where a pinhole was used as a spatial filter \cite{Peloso.2009}. These experiments were done without any mitigation of the effects of atmospheric turbulence. More recently, it was demonstrated that having only tip-tilt wavefront correction suffices for QKD in daylight through 53 km of atmosphere \cite{Liao.2017}. Soon after, the first demonstrations of daylight QKD with adaptive optics for higher-order wavefront correction have emerged \cite{Gong.2018,Gruneisen.2021}. All these employed a fixed spatial filter.

For optimal spatial filtering, the field of view (FOV) of the system needs to be reduced to roughly match the average angular extent of the signal focal spot. Finding the exact optimum, however, is far from straightforward due to the intricate interplay between received signal, noise, and the performance metric of the particular quantum communication protocol (e.g. secure key rate), and in fact even depends on the ever-changing link conditions. Moreover, there is an interplay between the different filtering degrees of freedom -- weaker filtering in one can be compensated by stronger filtering in another. Gruneisen~\textit{et~al.} comprehensively studied the concept of spatial filtering for space-to-Earth QKD with simulations \cite{Gruneisen.2016}. It can be seen from their results that the optimal spatial filter FOV, particularly in the case of tip-tilt-only correction, depends on the link characteristics, such as solar radiance, satellite altitude angle, and distance. Although not explicitly pointed out by Gruneisen~\textit{et~al.}, their plots indicate that having an adjustable spatial filter that could impose optimal filtering at all times would yield a considerable boost to the total generated key. We note that adaptive spatial filtering in their work refers to a fixed FOV filter with adaptive optics and not to a filter with an adaptive or adjustable FOV. Ko~\textit{et~al.} made the first attempts to experimentally address the question of optimal filtering and show an interplay between temporal and spatial degrees of freedom \cite{Ko.2018}. While they clearly find the optimum in the temporal domain, measurements with only two different fixed spatial filters did not offer much insight into optimal spatial filtering. Furthermore, their experiment was done over a short 275-m link, for which they reported a negligible effect of atmospheric turbulence.

Here, we introduce the concept of an adjustable spatial filter (SF) for applications in quantum communication, which can be adjusted to the optimum size under different link conditions. We built a prototype SF module that can seamlessly change the system FOV with $\upmu$rad precision. We integrated it into a free-space QKD system and experimentally demonstrated its use over a 1.7-km link in different link conditions. We report on our findings and compare optimal spatial filtering to simpler filtering with MM fibre coupling. Furthermore, we develop a theoretical model of our system and use it to investigate the potential of an adjustable spatial filter over a broader range of realistic link conditions.

\section{Methods}
\subsection{Experimental system}
\begin{figure}[h!]
    \centering\includegraphics[width=\textwidth]{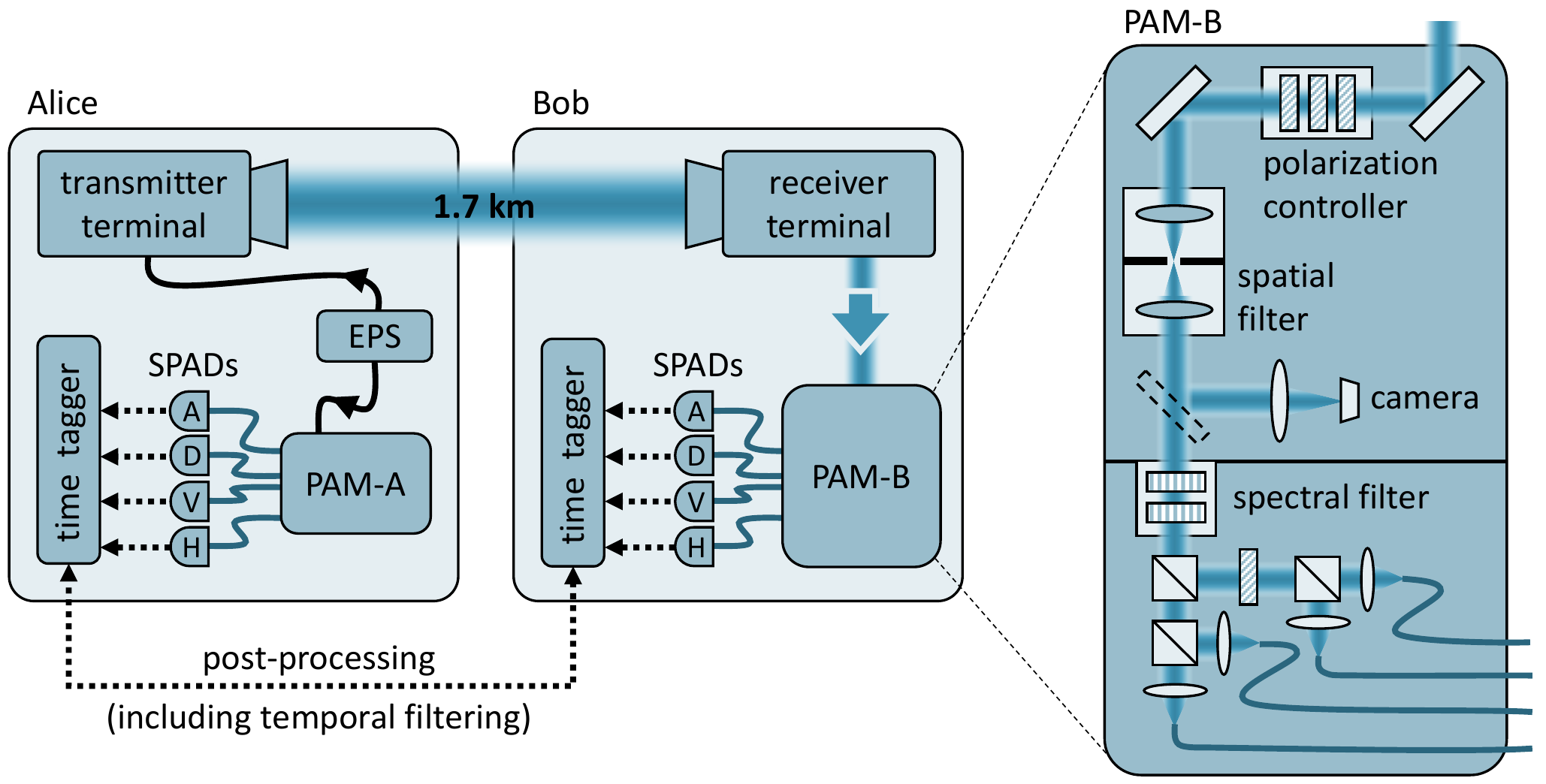}
    \caption{
        \label{fig:system_architecture}
        Schematic representation of the experimental system. Here, EPS is the entangled pair source, PAM-A is the polarization analysis module at Alice, PAM-B is the polarization analysis module at Bob, SPADs are single-photon avalanche diodes, and H, V, D, and A stand for horizontal, vertical, diagonal, and antidiagonal polarization, respectively.
    }
\end{figure}

To study the experimental impact of variable spatial filtering, we used the entanglement-based free-space quantum key distribution system reported in \cite{Krzic.25052022}, which is schematically shown in Fig.~\ref{fig:system_architecture}. The two communicating parties, Alice and Bob, are separated by a 1.7-km free-space channel, which is spanned by two identical telescopes with a primary aperture diameter of 200~mm \cite{Goy.3003202102042021}. To mitigate the effects of atmospheric turbulence, we use a 1064~nm beacon laser and tip-tilt correction at the receiver terminal. The quantum signal is generated at Alice with an entangled pair source (EPS), which generates polarization-entangled photon pairs at 810~nm by the means of spontaneous parametric down-conversion. One photon of each entangled pair is guided with a single-mode fibre to a local polarization analysis module (PAM). Here, polarization is measured in one of the two mutually unbiased bases, which without the loss of generality we call horizontal-vertical and diagonal-antidiagonal, and are chosen at random with 50:50 probability. For detection, we use commercial single-photon avalanche detectors (SPADs) with detection efficiency of $>$~60\% and a dark count rate of $<$~500~cps. Each detection event is given a precise arrival time by a rubidium-clock-driven time tagger. The other photon of each pair is emitted into free space, combined with the beacon beam, and sent to Bob, where it is similarly measured with a combination of a PAM, 4 detectors, and a time tagger. The data generated by both time taggers was stored and later processed offline, as explained below in Section \ref{sec:postprocessing}.

In addition to polarization analysis optics, Bob's PAM further incorporates dedicated spatial and spectral filters to prevent large amounts of background noise from reaching the detectors. The latter is realized using a stack of commercial interference filters, resulting in a passband with a full width at half maximum (FWHM) of 2.95~nm around 810~nm. For comparison, the EPS has a spectral bandwidth of 0.45~nm (FWHM), meaning that we do not utilize the full potential of spectral filtering here. The spatial filter with an adjustable FOV was developed specifically for the experiments here and is discussed in detail in Section \ref{sec:spatial_filter} below. Furthermore, Bob's PAM also includes a camera for imaging the focal plane of the spatial filter, which is used for alignment and characterization of the spatial filter and the incoming beam, and a waveplate-based polarization controller, which is used to align polarization frame of reference between Alice and Bob. When using the camera, a mirror is temporarily inserted into the beam path (dashed rectangle in Fig.~\ref{fig:system_architecture}). Alice's PAM is a much simpler version of Bob's, since it does not require dedicated filtering. This is due to the fact that it is directly connected to the EPS with a single-mode fibre and thus not exposed to the free-space link. Both PAMs were sealed, so that the light from the outside could enter Alice's PAM only through the input fibre and Bob's PAM only through a small aperture. Bob's PAM further incorporated an inner wall to prevent any stray light that was rejected by the filters to reach the polarization analysis part. 

Finally, note that after polarization analysis, the photons are coupled into MM fibres, which may also contribute to spatial filtering and is crucial for interpretation of our experimental results. Although having a dedicated spatial filtering module makes MM fibre coupling obsolete from the spatial filtering point of view, it is still practical for guiding light from the PAM to the detectors, allowing the PAM to be much more compact.

\subsection{Spatial filtering}
\label{sec:spatial_filter}
We developed a dedicated adjustable spatial filter (SF) that allowed us to vary the system FOV. It is realized with a lens-based telescope and a small opening in its focal plane, acting as a field stop, which was provided by Carl Zeiss Microscopy GmbH. It has a near-square shape and its side length is electronically adjustable between zero and about 400~$\upmu$m with a precision of a few microns. For our optical system, 1~$\upmu$m in the focal plane of the SF translates to about 0.344~$\upmu$rad in the receiver telescope primary aperture plane. In terms of the system FOV, the SF therefore has a dynamical range of about 0-140~$\upmu$rad and a precision of $\sim$1~$\upmu$rad. MM fibre coupling in PAM also acts as a field stop -- the nominal MM fibre core diameter expressed in the receiver telescope primary aperture is about 51.6~$\upmu$rad. In the following, unless explicitly stated otherwise, all the field values are given in angular units in the receiver telescope primary aperture plane.

\begin{figure}[h!]
\centering\includegraphics[width=\textwidth]{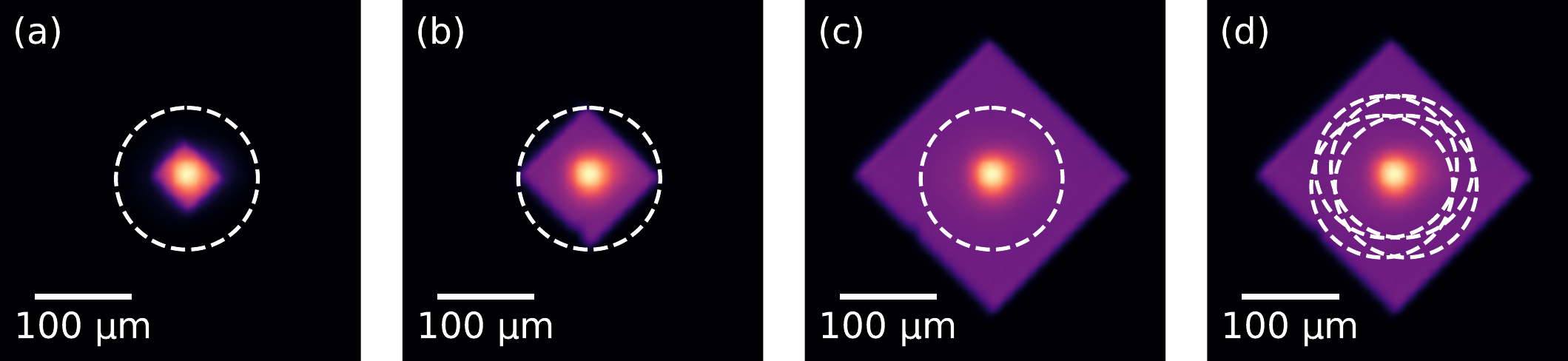}
\caption{
    \label{fig:SF_images}
    Average signal distribution, the adjustable field stop, and the multi-mode fibre cores, as seen in the focal plane of the spatial filter. The signal spot was measured while the tip-tilt correction was on. (a)-(c)~Signal with the field stop of various sizes and a schematic representation of ideal alignment of the fibres. (d)~Schematic representation of fibre misalignment.
}
\end{figure}

Aligning the SF field stop and the signal focal spot is of critical importance. For this purpose, we imaged the SF focal plane with the camera in Bob's PAM. To observe the field stop, we illuminated the SF with diffuse incoherent light. To observe the signal focal spot, we fully opened the SF and replaced the EPS by a laser of the same wavelength. Due to atmospheric turbulence, the short-term focal spot can be heavily distorted and its shape fluctuates randomly on millisecond timescales \cite{Andrews.2005}. However, what matters here is the long-term average of the spot, which dictates the mean signal throughput and can typically be well approximated with a Gaussian \cite{Andrews.2005}. Note that the size of the long-term focal spot varies with the strength of turbulence. While this effect is significantly reduced by the tip-tilt correction, the long-term size may still change considerably with time. At various occasions, we measured values ranging from 11.4~$\upmu$rad to 39.4~$\upmu$rad (FWHM). For comparison, the estimated FWHM of an ideal (diffraction-limited) spot for our system is 4.2~$\upmu$rad. An example long-term focal spot measurement (exposure time $>$20~s), superimposed with images of the SF field stop at different sizes, is shown in Fig.~\ref{fig:SF_images}. MM fibre cores are also drawn for reference (dashed circle). 

\begin{figure}[h!]
\centering
\includegraphics[width=\textwidth]{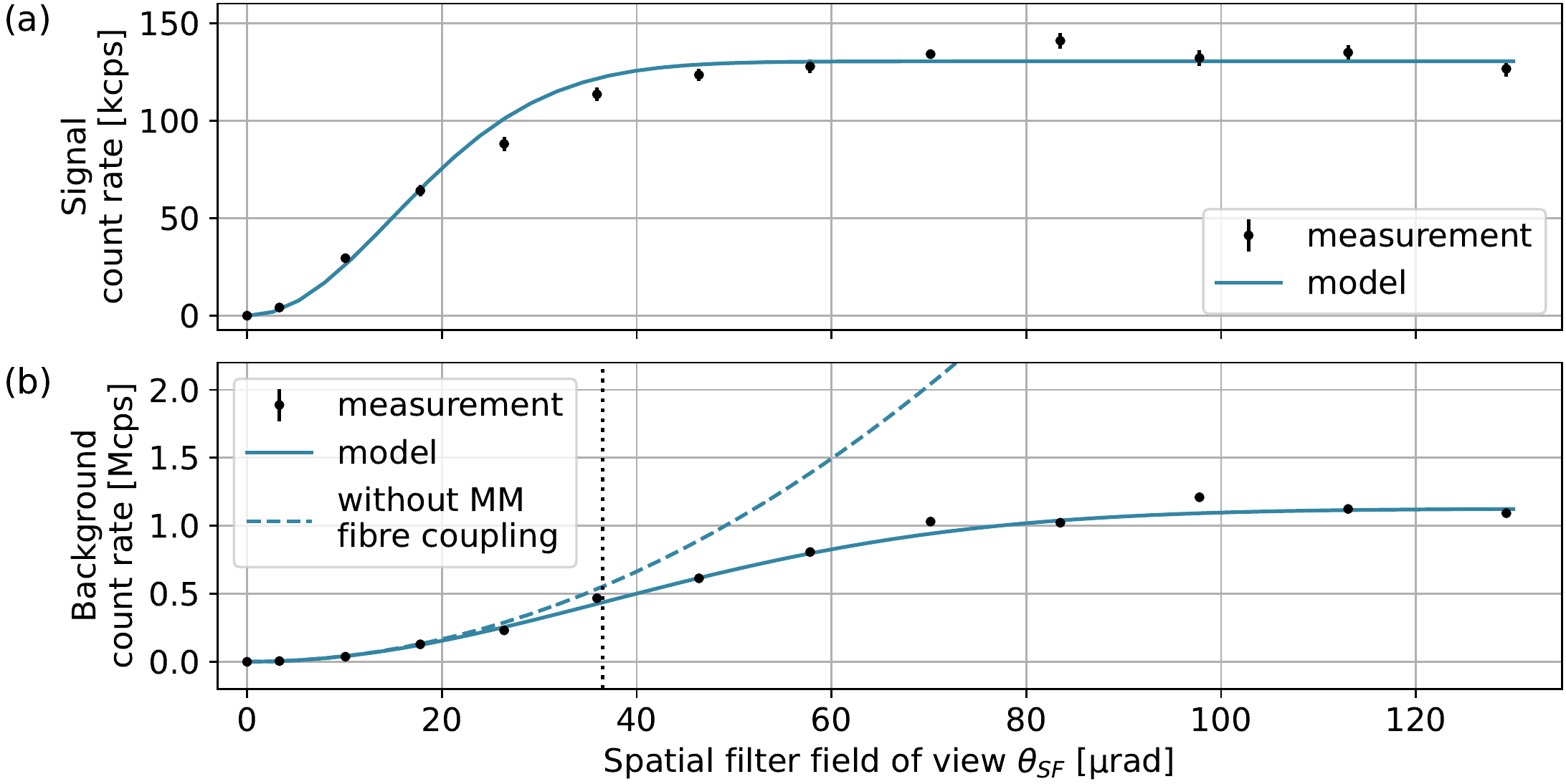}
\caption{
    \label{fig:signal_and_noise}
    Characterization of the spatial filter. (a)~Signal count rate dependence on the spatial filter field of view. (b)~Background noise count rate dependence on the spatial filter field of view. The dashed line is the corresponding estimate for the case if there was no multi-mode (MM) fibre coupling. The vertical dotted line indicates the point where the diagonal of the spatial filter matches the MM fibre core diameter.
}
\end{figure}

We then characterized the signal throughput dependence on the SF setting, which was done directly with the quantum signal from the EPS. At night, when the background noise was negligible, we scanned the SF size and recorded the total photon rate at Bob. The measurements are shown in Fig.~\ref{fig:signal_and_noise}(a). Since the mean focal spot can be well approximated with a Gaussian, we model the signal transmitted through the square-shaped field stop by a function of the form
\begin{equation}
\label{eq:signal}
    f_1(\theta_{\text{SF}};S_0,\Delta) = S_0 \,  \text{erf}^2\biggl[
        \frac{\sqrt{\text{ln}(2)}\theta_{\text{SF}}}{\Delta}
    \biggr] \, ,
\end{equation}
where $\text{erf}$ is the error function, $\theta_{\text{SF}}$ is the spatial filter FOV, corresponding to the side length of the field stop, $\Delta$ is the FWHM of the focal spot, and $S_0$ is the normalization factor, corresponding to the signal photon rate with fully open spatial filter. We can see that the chosen model (solid line) shows a good agreement with the measurements.

Similarly, we characterized the background noise dependence on the SF setting. These measurements were performed when the link was exposed to direct sunlight and the EPS turned off. One such set of measurements is shown in Fig.~\ref{fig:signal_and_noise}(b). The noise that is transmitted by the SF scales quadratically with $\theta_{\text{SF}}$ \cite{Gruneisen.2016}. At a certain value of $\theta_{\text{SF}}$, however, the MM fibres start limiting the system FOV, which breaks the quadratic dependence on $\theta_{\text{SF}}$ and the level of detected noise eventually saturates. In the case of perfect alignment of the MM fibres, all the four fibres would act as one and the noise would be given by an overlap integral of a square and a circle. Our measurements, however, showed a slower convergence towards the saturation than such a model would predict. We believe this is due to a slight misalignment of the MM fibres in our system, as schematically shown in Fig.~\ref{fig:SF_images}(d). The four fibres then effectively act as a single one with a smoother profile than a sharp circle, which results in a slower convergence of the overlap integral. We therefore looked for another function, which would still increase quadratically in the limit of small $\theta_{\text{SF}}$, while having a slower convergence towards saturation for larger $\theta_{\text{SF}}$. As it turns out, the measured data can be well described with a function of the form
\begin{equation}
\label{eq:background_noise}
    f_2(\theta_{\text{SF}}; B_0, \gamma) = B_0 \left(
        1 - \exp{\left[
            -\frac{\theta_{\text{SF}}^2}{2 {\gamma}^2}
        \right]}
    \right) \, ,
\end{equation}
where $B_0$ is the normalization factor and parameter $\gamma$ is the function turning point, i.e. the point at which the slope of the curve starts decreasing with $\theta_{\text{SF}}$.

The turning point $\gamma$ is expected to be a system constant that depends on the MM fibre coupling configuration. As $\theta_{\text{SF}}$ increases, the MM fibre coupling should start affecting the background rate when the diagonal of the SF field stop, which relates to $\sqrt{2}\theta_{\text{SF}}$, matches the diameter of the MM fibre core image in the SF field stop plane, as schematically shown in Fig.~\ref{fig:SF_images}(b). We measured the noise dependence on the SF setting on 9 different occasions throughout a day, each time finding a good agreement with Eq.~(\ref{eq:background_noise}). As predicted, all the fits resulted in very similar $\gamma$ values, having the mean and standard deviation of 36.8~$\upmu$rad and 1.4~$\upmu$rad, respectively. Moreover, $\sqrt{2}\gamma$ yields $52.0~\upmu\text{rad}$, which is indeed very close to the nominal MM fibre FOV of $51.6~\upmu\text{rad}$. This confirms $\gamma$ is a system constant and allows us to set it to a fixed value in our model. We are then left only with one free parameter to describe the amount of background noise, namely $B_0$, which represents the amount of noise in the case when the SF is completely open and the MM fibre coupling performs all the spatial filtering (i.e. MM fibre coupling limit). 

The corresponding fit of the data is shown in Fig.~\ref{fig:signal_and_noise}(b) (solid line). The point at which the SF field stop diagonal matches the MM fibre core diameter in the focal plane of the SF is also shown (vertical dotted line). We can clearly see that it matches the turning point of the fit function and marks the boundary between the two regimes: when spatial filtering is dominated by the adjustable spatial filter and when it is dominated by MM fibre coupling. To estimate background rates for the case if there was no MM fibre coupling (e.g. free-space detectors inside the PAM), we perform Taylor expansion of Eq.~(\ref{eq:background_noise}) and keep only the terms up to the quadratic one (dashed line), arriving at
\begin{equation}
\label{eq:background_noise_Taylor}
    \widetilde{f}_2(\theta_{\text{SF}}; B_0, \gamma) = B_0 \frac{\theta_{\text{SF}}^2}{2 {\gamma}^2} \, .
\end{equation}
This allows us to assess the role of MM fibres in our system as well as to generalize our results to a system without MM fibre coupling, as shown later in Section~\ref{sec:SF_effect_on_QKD}. This approximation will eventually break down, because the noise will not increase indefinitely with $\theta_{\text{SF}}$ but would saturate at the FOV limit imposed by the optical system before the SF. However, any practical detection system would most likely saturate well before that limit is reached. Note that the fibre-dependent $\gamma$ is part of the fibre-free model Eq.~(\ref{eq:background_noise_Taylor}) only because we defined $B_0$ in terms of noise collected by the fibres in the MM fibre coupling limit.

\subsection{Data acquisition and post-processing}
\label{sec:postprocessing}
For each data point, we recorded time tags of single-photon detection events at Alice and Bob for 8 seconds. The two sets of time tags were combined and processed offline, whereby differences in the local clocks were compensated for by exploiting the inherent time correlations of entangled photon pairs \cite{Spiess.2021}. We then extracted coincidences, which we define as events when a single-photon detection at one site happens within $\pm \tau/2$ of a detection time at the other site, where $\tau$ is called coincidence window \cite{Neumann.2021}. For our experiments, we chose $\tau = 800 \ \text{ps}$. Note that the extraction of coincidences essentially performs the function of temporal filtering.

With the coincidence rates for all detector combinations, we estimated the asymptotic secure key as 
\begin{equation}
\label{eq:SKR}
    R = C_{\text{sift}} \left[1 - f \text{H}_2(E_{\text{HV}}) - \text{H}_2(E_{\text{DA}}) \right] \,
\end{equation}
where $C_{\text{sift}}$ is the total coincidence rate after sifting, i.e. coincidences in the same measurement basis at Alice and Bob, $f$ is the error correction efficiency for which we for simplicity take a constant value of 1.22 \cite{Waks.2002}, $\text{H}_2(x)$ is the binary entropy function defined as $\text{H}_2(x) = -x \log_2(x) - (1 - x)\log_2(1 - x)$, and $E_{\text{HV}}$ and $E_{\text{DA}}$ are the measured error rates in horizontal-vertical and diagonal-antidiagonal basis, respectively \cite{Neumann.2021}.

\subsection{Model for key rate dependence on spatial filtering}
\label{sec:QKD_model}
We now integrate the spatial filtering model above into the QKD model from \cite{Neumann.2021}. Assuming the error rate to be the same in both bases, it is given by
\begin{equation}
    \label{eq:error_rate}
    E = E_{\text{HV}} = E_{\text{DA}} = \frac{{\eta}^{c} C^\text{t} E^{\text{pol}} + \frac{1}{2} C^{\text{acc}}} {{\eta}^{c} C^\text{t} + C^{\text{acc}}} \, ,
\end{equation}
where $C^\text{t}$ is the true coincidence rate, $C^{\text{acc}}$ is the accidental coincidence rate, ${\eta}^{c}$ is the coincidence-window-dependent detection efficiency, and $E^{\text{pol}}$ is the baseline error rate due to polarization mismatch. Furthermore,
\begin{equation}
    \label{eq:true_coincidences}
    C^\text{t} = \eta_{\text{A}} S_\text{B}^\text{t} \, ,
\end{equation}
\begin{equation}
    C^{\text{acc}} = \frac{\left(1 - \text{e}^{-S_\text{A}^\text{m}\tau}\right) \left(1 -  \text{e}^{-S_\text{B}^\text{m}\tau}\right)}{\tau} \, ,
\end{equation}
and
\begin{equation}
    {\eta}^{c} = \text{erf} \left[ \sqrt{\text{ln}(2)} \frac{\tau}{t_{\Delta}} \right] \, ,
\end{equation}
where $\eta_{\text{A}}$ is the total single photon efficiency in Alice's channel, $S_\text{B}^\text{t}$ is the true (signal) photon rate at Bob, $S_\text{A}^\text{m}$ and $S_\text{B}^\text{m}$ are the total measured photon rates at Alice and Bob, respectively, and $ t_{\Delta}$ is the FWHM of the total system timing jitter. $C^\text{t}$ is related to the measured coincidence rate $C^\text{m}$ through
\begin{equation}
    C^\text{m} = {\eta}^{c}C^\text{t} + C^{\text{acc}} \, .
\end{equation}

We adapt the model to our experimental system by expressing $S_\text{B}^\text{t}$ with Eq.~(\ref{eq:signal}) and defining
\begin{equation}
\label{eq:total_Bob_count}
    S_\text{B}^\text{m} = S_\text{B}^\text{t} + S_\text{B}^\text{b} + S_\text{B}^\text{d} \, ,
\end{equation}
where $S_\text{B}^\text{d}$ is the dark count rate and $S_\text{B}^\text{b}$ is the background photon count rate at Bob according to Eq.~(\ref{eq:background_noise}) or Eq.~(\ref{eq:background_noise_Taylor}), depending on whether the MM fibres are considered or not, using the empirically determined $\gamma$.

For the experiments here, most of the remaining model parameters can be set to a fixed value. $S_\text{B}^\text{d} = 761 \ \text{cps}$ and $t_{\Delta} = 710 \ \text{ps}$ are system constants and were measured in advance. $\eta_{\text{A}}$ consists of intrinsic EPS efficiency, the fibre-only channel efficiency, and detector efficiency, all of which are sufficiently stable to be considered constant over the course of our experiments. We estimated $\eta_{\text{A}} = 12.2\%$, which was done by measuring $C^\text{m}$, $S_\text{A}^\text{m}$, and $S_\text{B}^\text{m}$ at night, when $S_\text{B}^\text{b} \approx 0$, and using Eqs.~(\ref{eq:true_coincidences})-(\ref{eq:total_Bob_count}). Similarly, $S_\text{A}^\text{m}$ was near-constant at 1.4~Mcps. We determined $E^{\text{pol}}$ as
\begin{equation}
    E^{\text{pol}} = \frac{C^{\text{wrong}} - \frac{1}{4}C^{\text{acc}}}{C^{\text{sift}}} \, , 
\end{equation}
where 
\begin{equation}
    C^{\text{right}} = C^{\text{HV}}+C^{\text{VH}}+C^{\text{DA}}+C^{\text{AD}} \, , 
\end{equation}
\begin{equation}
    C^{\text{wrong}} = C^{\text{HH}}+C^{\text{VV}}+C^{\text{DD}}+C^{\text{AA}} \, , 
\end{equation}
\begin{equation}
    C^{\text{sift}} = C^{\text{right}} + C^{\text{wrong}} \, , 
\end{equation}
and $C^{\text{ij}}$ is the coincidence rate between i-th channel at Alice and j-th channel at Bob. We estimated $E^{\text{pol}} = 2.6\%$, which we also assume to be constant throughout the day of experiments.

Finally, using Eqs.~(\ref{eq:signal})-(\ref{eq:background_noise_Taylor})~and~(\ref{eq:error_rate})-(\ref{eq:total_Bob_count}) with Eq.~(\ref{eq:SKR}), together with the fixed parameters from above, we arrive at a secure key rate model $R(\theta_{\text{SF}}; S_0, \Delta, B_0)$ that is a function of the SF setting and has three link-conditions-dependent free parameters. In Section~\ref{sec:results}, we use this model for fitting the experimentally determined key rates and for extrapolation to other potential link conditions. Note, that $B_0$ could be independently characterized at the time of each experiment by turning the EPS off and measuring and fitting the noise-only dependence on $\theta_{\text{SF}}$, as in Fig.~\ref{fig:signal_and_noise}(b). On the other hand, removing the background noise to measure only signal was not possible. Moreover, with the quantum signal, it was also not possible to directly measure the focal spot. However, fitting the model to the measurements not only allowed us to interpolate and extrapolate the data but also provided us with an estimate for $\Delta$ and $S_0$, which offered further insights into the link conditions at the time of the experiments.

\section{Results}
\label{sec:results}

\subsection{Experimental demonstration of adjustable spatial filtering}
\label{sec:SF_effect_on_QKD}
\begin{figure}[h!]
\centering\includegraphics[width=11cm]{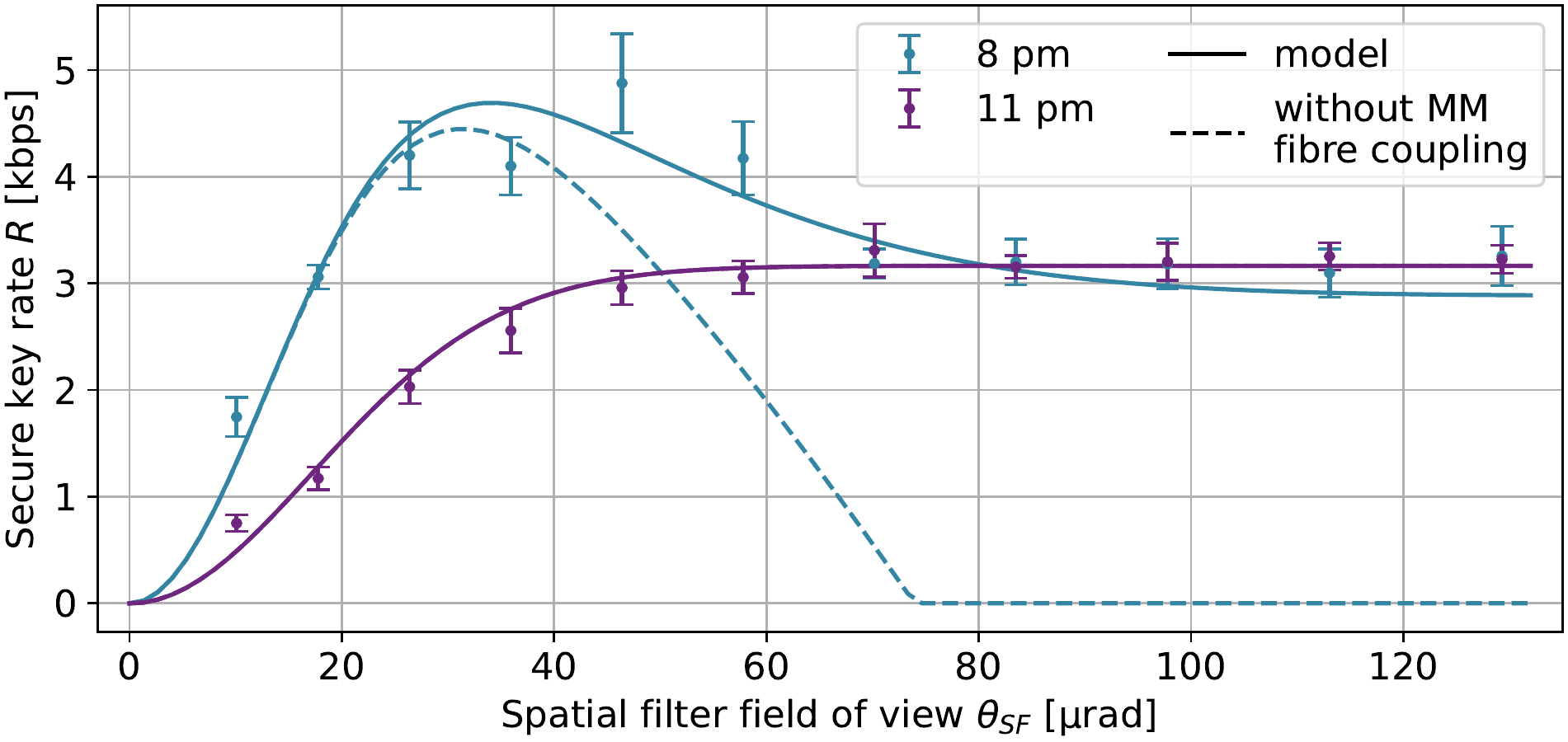}
\caption{
    \label{fig:SKR_vs_FOV}
    Experimentally determined asymptotic secure key rate dependence on the spatial filter field of view (points) and the corresponding fit (solid line) for two experiments with different link conditions. The dashed lines show the corresponding estimates for the case if there was no multi-mode (MM) fibre coupling at the receiver.
}
\end{figure}

We now experimentally demonstrate the use of the adjustable spatial filter for quantum key distribution. On 15 June 2022, we performed two QKD experiments. The first experiment was done at about 8~pm, right before sunset, when the Sun was still fully above the horizon and the link was exposed to direct sunlight. At this time, we recorded the highest noise rates during the entire day. This is most likely due to the fact that Bob's receiver was facing north-west, meaning that the angle between the receiver telescope orientation and the Sun was the smallest then. The second experiment was done at about 11~pm, long enough after the sunset, so that the background noise was at a negligible level. Each experiment consisted of a simultaneous recording of time tags at both sites for a range of SF settings, which lasted several minutes in total. Right after, we also measured the background-only dependence on the SF setting by turning the EPS off. During the measurements within each experiment, we assume the link conditions, i.e. the amount of sunlight and the strength of atmospheric turbulence, did not considerably change. The recorded time tags were later synchronized and post-processed to calculate the secret key rates, as explained in Section~\ref{sec:postprocessing}. The results are shown in Fig.~\ref{fig:SKR_vs_FOV}. From the background-only measurements, we extracted the background noise normalization factor $B_0$, as explained in Section~\ref{sec:spatial_filter}. Using this $B_0$, we then fit the experimentally determined key rates with the model $R(\theta_{\text{SF}}; S_0, \Delta, B_0)$ described in Section~\ref{sec:QKD_model} (solid lines), thus extracting the signal spot size $\Delta$ and the signal normalization factor $S_0$. The corresponding estimated secure key rate in the absence of the MM fibres is also shown (dashed lines). 

The extracted parameters are given in Table~\ref{tab:SKR_fit_parameters}. Since $\Delta$ is mostly influenced by the residual wavefront aberrations uncorrected by the tip-tilt system, it may give us an idea about the atmospheric turbulence strength during the experiments – the larger the spot size, the stronger the turbulence. The extracted values therefore indicate that the turbulence was weaker right before sunset than later at night. This is not surprising, since the surface and air temperatures around sunset are typically nearly identical, resulting in lower wind speeds and weaker turbulence at that time \cite{Andrews.2005}. Turbulence strength not only influences the focal spot size but also link efficiency by spreading the beam size at the receiver telescope aperture, in turn reducing receiver collection efficiency. Moreover, stronger turbulence makes the link alignment more difficult, thus limiting the attainable alignment precision and further lowering the link efficiency. This explains why the extracted $S_0$ was larger at 8~pm, despite the EPS output being almost the same during both experiments. The difference in $S_0$ was thus mostly due to the difference in link efficiency. Consequently, higher key rates could be achieved in direct sunlight than in the night. However, if we had the same turbulence at 11~pm as at 8~pm, the key rate would saturate at a value larger than 6~kbps.

\begin{table}[ht]
    \caption{Extracted model parameters for the experiments shown in Fig.~\ref{fig:SKR_vs_FOV}.} 
    \label{tab:SKR_fit_parameters}
    \begin{center}       
    \begin{tabular}{|l|l|l|l|} 
    \hline
    \rule[-1ex]{0pt}{3.5ex}   & $B_0$ [Mcps] & $\Delta$ [$\upmu$rad] & $S_0$ [kcps] \\
    \hline
    \rule[-1ex]{0pt}{3.5ex}  8 pm & 1.59 $\pm$ 0.03 & 18.5 $\pm$ 0.7 & 237 $\pm$ 5  \\
    \hline
    \rule[-1ex]{0pt}{3.5ex}  11 pm & $\sim$ 0 & 23.0 $\pm$ 1.2 & 117 $\pm$ 3 \\
    \hline
    \end{tabular}
    \end{center}
\end{table}

The experiments clearly show that under certain conditions, spatial filtering by the means of MM fibre coupling alone can be sufficient for QKD. However, further reducing the FOV to the optimal value still results in a considerable increase of the achievable secure key rate in daytime. The improvement is even more pronounced in the case without MM fibre coupling. In nighttime, such strong levels of spatial filtering are not needed anymore, therefore fully opening the SF field stop becomes the optimal strategy, with or without MM fibre coupling (the two curves overlap). These results already indicate that the strategy to be actively adjusting the system FOV to its optimum, as the link conditions change, has the potential to improve the total secure key output of the system. The two experiments, however, do not cover a wide range of possible link conditions. Moreover, since we cannot change the link-conditions-dependent parameters at will, we cannot experimentally observe the effect of each one of them individually over the actual free-space link. Additionally optimizing also temporal filtering, which was not done here, would improve the key rates and might further affect the optimal spatial filtering configuration. In the next section, we thus use the developed model to extrapolate our experimental results to a broader range of possible link conditions, changing one parameter at a time, while also optimizing the coincidence window for each data point.

\subsection{Extrapolation to different link conditions}
To extend the study to a wider range of terrestrial free-space link conditions, we take the model and parameters extracted from the 8 pm experiment and vary one parameter at a time. We limit ourselves to the more general case without MM fibre coupling. Furthermore, we numerically optimize coincidence window for each data point. We thus extrapolate the key rate dependence on the spatial filter FOV to a set of evenly spaced values over the following intervals: $B_0 \in [0, 1.59]$~Mcps, $\Delta \in [11.4, 39.4]$~$\upmu$rad, and $S_0 \in [0, 237]$~kcps. The bounds correspond to the largest and the smallest values we have recorded with our system in various experiments over several months.

\begin{figure}[h!]
\centering\includegraphics[width=11.5cm]{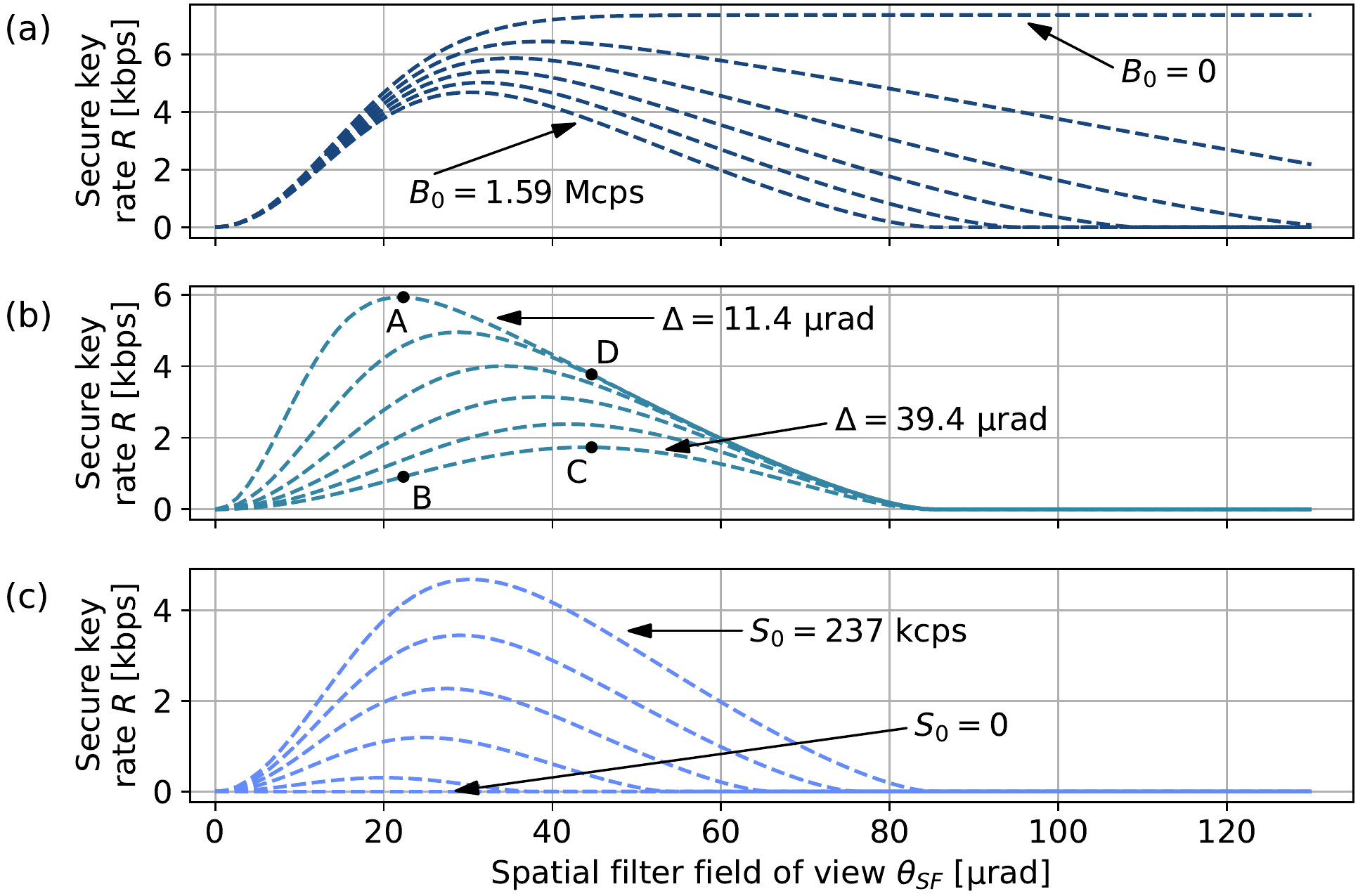}
\caption{
    \label{fig:SKR_parameter_scan}
    Extrapolation of the model extracted from the 8~pm experiment to different link conditions, assuming no fibre coupling and optimal coincidence window. In each plot, one of the link-conditions-dependent parameters is varied: (a)~background noise count rate, (b)~mean focal spot size, and (c)~signal count rate.
}
\end{figure}

The results are shown in Fig.~\ref{fig:SKR_parameter_scan}. We may clearly see that the largest influence on the optimal FOV is due to the change of $\Delta$. To quantitatively benchmark the potential of the adjustable spatial filter to improve the overall output of the system, let us now consider a hypothetical scenario, where only $\Delta$ would change while $B_0$ and $S_0$ would stay constant. Suppose we would be experiencing weak turbulence, corresponding to the smallest $\Delta$ considered in our extrapolations, and the system would be operating with the optimal FOV (point~A). The corresponding key rate would be 5.9~kbps. If the conditions would then worsen to the point where $\Delta$ would be the largest considered here, the key rate would drop to 0.85~kbps (point~B). With the adjustable spatial filter, we could then optimize the spatial filter FOV and increase the key rate to 1.7~kbps (point~C), which is a 2-fold improvement. Should the conditions improve again to the starting ones, the key rate would increase to 3.8~kbps (point~D). We could then again optimize the FOV and increase the key rate back to 5.9~kbps (point~A), which is a 1.5-fold improvement. For a system like ours, an adjustable spatial filter therefore has the potential to improve the overall key output by a substantial amount.

$B_0$ and $S_0$ also affect the optimal FOV, but to a smaller extent. We may also note that worsening of $B_0$ and $S_0$ (i.e. increase and decrease, respectively) shifts the optimal FOV to smaller values, while worsening of $\Delta$ (increase) shifts it to larger values. $S_0$, on the other hand, has a strong influence on the secure key rate. This explains why we could achieve higher key rates at 8~pm in direct sunlight than at 11~pm, when the background noise was negligible. In real QKD, an interplay between these effects is therefore taking place.


\section{Conclusion}
We developed an adjustable spatial filter for quantum communication in daylight and introduced the idea of active optimization of the system field of view with the ever-changing link conditions. We demonstrated the filter in a real QKD system with tip-tilt-only wavefront correction over a 1.7-km free-space channel and showed it has the potential to substantially improve the total secure key rate output.

The exact factor of improvement, however, heavily depends on the free-space channel conditions, in particular turbulence strength and the amount of daylight noise. Furthermore, it depends on the underlying system, such as the level of filtering in other degrees of freedom and the level of compensation of atmospheric turbulence. For systems without any wavefront correction at the receiver, an adjustable spatial filter would be of even higher practical relevance, since the focal spot size would vary to a much greater extent with the varying turbulence strength. On the other hand, if the system incorporated full adaptive optics, the benefit of having it adjustable would be much less pronounced, since the focal spot size would not significantly change with link conditions, as it is evident from the simulations in \cite{Gruneisen.2016}. High-performance adaptive optics would also enable efficient single-mode fibre coupling, which is the extremest form of spatial filtering. However, our work shows that intermediate levels of spatial filtering can also offer considerable benefits for quantum communication in scenarios where only low-order wavefront correction is desirable or possible.

While the experiments and extrapolations in this work were done for a terrestrial free-space link, the approach can be readily extended to satellite links, where the optimal field of view would further depend on the satellite orbit and change with the satellite altitude angle \cite{Gruneisen.2016}. This means that an adjustable spatial filter in an optical ground station would allow for optimal performance with a single system, not only at any time of the day but also for a link with any satellite. Finally, we stress that although we demonstrated the adjustable spatial filter with polarization-based QKD, the approach is suitable for improving performance of other encodings and protocols in free-space optical communication -- quantum as well as classical.

\section*{Funding} 
German Federal Ministry of Education and Research - BMBF (16KIS1263K); European Space Agency - ESA (4000125842/18/NL/MH/mg).


\section*{Acknowledgments} 
We thank Herbert Gross for fruitful discussions and help with the spatial filter design; Mirko Liedtke and Carl Zeiss Microscopy GmbH for providing the adjustable field stop; Sakshi Sharma for help with the entangled pair source; Matthias Goy and Daniel Heinig for experimental support; Stadtwerke Jena for giving us access to their rooftop for experiments. This research was conducted within the scope of the project QuNET, funded by the German Federal Ministry of Education and Research (BMBF) in the context of the federal government’s research framework in IT-security “Digital. Secure. Sovereign.” AK and CS are part of the Max Planck School of Photonics supported by the BMBF, the Max Planck Society, and the Fraunhofer Society. AK is supported by ESA through the Networking/Partnering Initiative.



\printbibliography

@incollection{Krenn.2016,
 author = {Krenn, Mario and Malik, Mehul and Scheidl, Thomas and Ursin, Rupert and Zeilinger, Anton},
 title = {Quantum Communication with Photons},
 pages = {455--482},
 volume = {84},
 publisher = {{Springer International Publishing}},
 isbn = {978-3-319-31902-5},
 editor = {Al-Amri, Mohammad D. and El-Gomati, Mohamed and Zubairy, M. Suhail},
 booktitle = {Optics in Our Time},
 year = {2016},
 address = {Cham},
 doi = {10.1007/978-3-319-31903-2{\_}18}
}

@article{Zhang.2018,
 author = {Zhang, Qiang and Xu, Feihu and Chen, Yu-Ao and Peng, Cheng-Zhi and Pan, Jian-Wei},
 year = {2018},
 title = {Large scale quantum key distribution: challenges and solutions Invited},
 pages = {24260--24273},
 volume = {26},
 number = {18},
 journal = {Optics express},
 doi = {10.1364/OE.26.024260}
}

@article{Gruneisen.2015,
 author = {Gruneisen, Mark T. and Flanagan, Michael B. and Sickmiller, Brett A. and Black, James P. and Stoltenberg, Kurt E. and Duchane, Alexander W.},
 year = {2015},
 title = {Modeling daytime sky access for a satellite quantum key distribution downlink},
 pages = {23924},
 volume = {23},
 number = {18},
 journal = {Optics express},
 doi = {10.1364/OE.23.023924}
}

@book{Andrews.2005,
 author = {Andrews, Larry C. and Phillips, Ronald L.},
 year = {2005},
 title = {Laser beam propagation through random media},
 address = {Bellingham, Wash.},
 edition = {2nd ed.},
 publisher = {{SPIE Press}},
 isbn = {0-8194-5948-8}
}

@article{Hughes.2000,
 author = {Hughes, Richard J. and Buttler, William T. and Kwiat, Paul G. and Lamoreaux, Steve K. and Morgan, George L. and Nordholt, Jane E. and Peterson, C. Glen},
 year = {2000},
 title = {Free-space quantum key distribution in daylight},
 pages = {549--562},
 volume = {47},
 number = {2-3},
 issn = {0950-0340},
 journal = {Journal of Modern Optics},
 doi = {10.1080/09500340008244059}
}

@article{Peloso.2009,
 author = {Peloso, Matthew P. and Gerhardt, Ilja and Ho, Caleb and Lamas-Linares, Ant{\'i}a and Kurtsiefer, Christian},
 year = {2009},
 title = {Daylight operation of a free space, entanglement-based quantum key distribution system},
 pages = {045007},
 volume = {11},
 number = {4},
 issn = {1367-2630},
 journal = {New Journal of Physics},
 doi = {10.1088/1367-2630/11/4/045007}
}

@article{Liao.2017,
 author = {Liao, Sheng-Kai and Yong, Hai-Lin and Liu, Chang and Shentu, Guo-Liang and Li, Dong-Dong and Lin, Jin and Dai, Hui and Zhao, Shuang-Qiang and Li, Bo and Guan, Jian-Yu and Chen, Wei and Gong, Yun-Hong and Li, Yang and Lin, Ze-Hong and Pan, Ge-Sheng and Pelc, Jason S. and Fejer, M. M. and Zhang, Wen-Zhuo and Liu, Wei-Yue and Yin, Juan and Ren, Ji-Gang and Wang, Xiang-Bin and Zhang, Qiang and Peng, Cheng-Zhi and Pan, Jian-Wei},
 year = {2017},
 title = {Long-distance free-space quantum key distribution in daylight towards inter-satellite communication},
 pages = {509--513},
 volume = {11},
 number = {8},
 issn = {1749-4885},
 journal = {Nature Photonics},
 doi = {10.1038/NPHOTON.2017.116}
}

@article{Gong.2018,
 author = {Gong, Yun-Hong and Yang, Kui-Xing and Yong, Hai-Lin and Guan, Jian-Yu and Shentu, Guo-Liang and Liu, Chang and Li, Feng-Zhi and Cao, Yuan and Yin, Juan and Liao, Sheng-Kai and Ren, Ji-Gang and Zhang, Qiang and Peng, Cheng-Zhi and Pan, Jian-Wei},
 year = {2018},
 title = {Free-space quantum key distribution in urban daylight with the SPGD algorithm control of a deformable mirror},
 pages = {18897--18905},
 volume = {26},
 number = {15},
 journal = {Optics express},
 doi = {10.1364/OE.26.018897}
}

@article{Gruneisen.2021,
 author = {Gruneisen, Mark T. and Eickhoff, Mark L. and Newey, Scott C. and Stoltenberg, Kurt E. and Morris, Jeffery F. and Bareian, Michael and Harris, Mark A. and Oesch, Denis W. and Oliker, Michael D. and Flanagan, Michael B. and Kay, Brian T. and Schiller, Johnathan D. and Lanning, R. Nicholas},
 year = {2021},
 title = {Adaptive-Optics-Enabled Quantum Communication: A Technique for Daytime Space-To-Earth Links},
 pages = {126111},
 volume = {16},
 number = {1},
 journal = {Physical Review Applied},
 doi = {10.1103/PhysRevApplied.16.014067}
}

@article{Gruneisen.2016,
 author = {Gruneisen, Mark T. and Sickmiller, Brett A. and Flanagan, Michael B. and Black, James P. and Stoltenberg, Kurt E. and Duchane, Alexander W.},
 year = {2016},
 title = {Adaptive spatial filtering of daytime sky noise in a satellite quantum key distribution downlink receiver},
 pages = {026104},
 volume = {55},
 number = {2},
 issn = {0091-3286},
 journal = {Optical Engineering},
 doi = {10.1117/1.OE.55.2.026104}
}

@article{Ko.2018,
 author = {Ko, Heasin and Kim, Kap-Joong and Choe, Joong-Seon and Choi, Byung-Seok and Kim, Jong-Hoi and Baek, Yongsoon and Youn, Chun Ju},
 year = {2018},
 title = {Experimental filtering effect on the daylight operation of a free-space quantum key distribution},
 pages = {15315},
 volume = {8},
 number = {1},
 journal = {Scientific reports},
 doi = {10.1038/s41598-018-33699-y}
}

@inproceedings{Goy.3003202102042021,
 author = {Goy, Matthias and Berlich, Ren{\'e} and Kr{\v{z}}i{\v{c}}, Andrej and Riel{\"a}nder, Daniel and Kopf, Teresa and Sharma, Sakshi and Steinlechner, Fabian O.},
 title = {High performance optical free-space links for quantum communications},
 url = {https://www.spiedigitallibrary.org/conference-proceedings-of-spie/11852/2599163/High-performance-optical-free-space-links-for-quantum-communications/10.1117/12.2599163.full},
 pages = {18},
 publisher = {SPIE},
 isbn = {9781510645486},
 editor = {Sodnik, Zoran and Cugny, Bruno and Karafolas, Nikos},
 booktitle = {International Conference on Space Optics --- ICSO 2020},
 year = {2021},
 doi = {10.1117/12.2599163}
}

@article{Neumann.2021,
 author = {Neumann, Sebastian Philipp and Scheidl, Thomas and Selimovic, Mirela and Pivoluska, Matej and Liu, Bo and Bohmann, Martin and Ursin, Rupert},
 year = {2021},
 title = {Model for optimizing quantum key distribution with continuous-wave pumped entangled-photon sources},
 pages = {226},
 volume = {104},
 number = {2},
 issn = {1050-2947},
 journal = {Physical Review A},
 doi = {10.1103/PhysRevA.104.022406}
}

@article{Krzic.25052022,
 author = {Kr{\v{z}}i{\v{c}}, Andrej and Sharma, Sakshi and Spiess, Christopher and Chandrashekara, Uday and T{\"o}pfer, Sebastian and Sauer, Gregor and Campo, Luis Javier Gonz{\'a}lez-Mart{\'i}n del and Kopf, Teresa and Petscharnig, Stefan and Grafenauer, Thomas and Lieger, Roland and {\"O}mer, Bernhard and Pacher, Christoph and Berlich, Ren{\'e} and Peschel, Thomas and Damm, Christoph and Risse, Stefan and Goy, Matthias and Riel{\"a}nder, Daniel and T{\"u}nnermann, Andreas and Steinlechner, Fabian},
 date = {2022},
 title = {Metropolitan free-space quantum networks},
 note = {preprint at \url{https://arxiv.org/abs/2205.12862}},
}

@article{Spiess.2021,
 author = {Spiess, Christopher and T{\"o}pfer, Sebastian and Sharma, Sakshi and Kr{\v{z}}i{\v{c}}, Andrej and Ponce, Meritxell Cabrejo and Chandrashekara, Uday and D{\"o}ll, Nico Lennart and Riel{\"a}nder, Daniel and Steinlechner, Fabian},
 date = {2021},
 title = {Clock synchronization with correlated photons},
 note = {preprint at \url{https://arxiv.org/abs/2108.13466}},
}

@article{Pirandola.2020,
 author = {Pirandola, S. and Andersen, U. L. and Banchi, L. and Berta, M. and Bunandar, D. and Colbeck, R. and Englund, D. and Gehring, T. and Lupo, C. and Ottaviani, C. and Pereira, J. L. and Razavi, M. and {Shamsul Shaari}, J. and Tomamichel, M. and Usenko, V. C. and Vallone, G. and Villoresi, P. and Wallden, P.},
 year = {2020},
 title = {Advances in quantum cryptography},
 pages = {1012},
 volume = {12},
 number = {4},
 journal = {Advances in Optics and Photonics},
 doi = {10.1364/AOP.361502}
}

@article{Waks.2002,
 author = {Waks, Edo and Zeevi, Assaf and Yamamoto, Yoshihisa},
 year = {2002},
 title = {Security of quantum key distribution with entangled photons against individual attacks},
 pages = {3121},
 volume = {65},
 number = {5},
 issn = {1050-2947},
 journal = {Physical Review A},
 doi = {10.1103/PhysRevA.65.052310}
}

\end{document}